\documentclass[aps,graphicx,6pt,showkeys,showpacs,twocolumn]{revtex4}
\usepackage{amsmath}
\usepackage{amscd}
\usepackage{graphicx}
\usepackage{subfigure}
\usepackage{appendix}
\usepackage{amsfonts}
\usepackage{multirow}
\usepackage{bm}

\begin{document}

\title[Short Title]{Optimal shortcut approach based on an easily obtained intermediate Hamiltonian}

\author{Ye-Hong Chen$^{1,2}$}
\author{Zhi-Cheng Shi$^{1,2}$}
\author{Jie Song$^{3}$}
\author{Yan Xia$^{1,2,}$\footnote{E-mail: xia-208@163.com}}
\author{Shi-Biao Zheng$^{1,2}$}

\affiliation{$^{1}$Department of Physics, Fuzhou University, Fuzhou 350116, China\\
             $^{2}$Fujian Key Laboratory of Quantum Information and Quantum Optics (Fuzhou University), Fuzhou 350116, China\\
             $^{3}$Department of Physics, Harbin Institute of Technology, Harbin 150001, China}


\begin{abstract}
  We present a general approach to speed up the adiabatic process
  without adding the traditional counterdiabatic driving (CD) Hamiltonian.
  The strategy is to design an easy-to-get intermediate Hamiltonian
  to connect the original Hamiltonian and final transitionless Hamiltonian.
  With final transitionless Hamiltonian, the same target can be achieved
  as in the adiabatic process governed by the original Hamiltonian, but in a shorter time.
  We apply the present approach to a three-level system, and the result shows that
  the final transitionless Hamiltonian usually has the same structure as the
  original Hamiltonian but with different time-dependent coefficients, allowing speedup to be achieved in a much easier way compared to previous methods.
\end{abstract}


\maketitle
\section{Introduction}
Coherent control of the quantum state is a critical element for various quantum technologies such as high-precision
measurement \cite{Rmp781297}, coherent manipulation of atom and
molecular systems \cite{Rmp7953}, and quantum information processing \cite{WVB08,Prl95080502}.
Approaches usually used to realize coherent control, however,
are subject to some deficiencies. For example, the coherent Rabi oscillation approach
is fast but very sensitive to parameter fluctuations.
On the contrary, adiabatic approach (including its variances) is
insensitive to the noises of the
applied drivings, but the adiabatic condition requires a long time.
In recent years, various versions of the approach named ``shortcuts to adiabaticity'' (STA)
\cite{Prl105123003,ETSISMGMMACDGOARXCJGMAmop13,Prl111100502,Prl109100403} have been proposed to
speed up the adiabatic process while hold
the advantages of robustness against parameter fluctuations by relaxing the adiabatic requirements,
including, e.g., transitionless driving algorithm (TDA) \cite{Jpca1079937,Jpa42365303,Oc13948,Epjst224189,
Pra82033430,Epl9323001},
inverse engineering based on invariants \cite{Pra86033405,Pra89053408,Pra89043408}, and so on
\cite{Jpb43085509,Pra84031606Epl9660005,arXiv160105551,Njp16015025,Prl116230503}.
The basic idea of these methods is to design a specific supplementary driving
to eliminate the undesired non-adiabatic transitions which
practically exist in the original adiabatic process.
With this, one can remove the relevant adiabatic condition, and thus the
transitionless evolution of the instantaneous eigenvectors
could be driven fast.

However, the designed supplementary Hamiltonian (the CD Hamiltonian)
usually introduces some undesired off-diagonal terms to the original
Hamiltonian which might cause problems in realizing the completed
Hamiltonian in practice. For instance, in a recent
experimental realization of speeding up a stimulated Raman adiabatic
passage (STIRAP) with cold atoms  \cite{Natc712479}, these undesired terms are suppressed by reducing the dynamics of a three-level system to that of a two-level system under the large detuning condition, which limits the operation time. To overcome the
problem caused by the CD Hamiltonian, several ingenious methods have
been proposed in recent years
\cite{Prl109100403,Pra86033405,Pra89053408,Pra89043408,Prl116230503,Pra93052109}.
Generally speaking, the basic idea of the previous approaches is constructing another path for the system so that one can drive the system to evolve along the new path
to attain the same target state as the desired adiabatic path. The condition is that the new path should be ensured
to coincide with the desired adiabatic path at the initial time ($t_{i}$) and the final time $(t_{f})$.
Specifically, the new paths are: (i) the eigenvectors of the invariant operators,
(ii) the eigenvectors of the $n$th iteration Hamiltonian, (iii) the dressed states,
in approaches of: (i) invariants-based shortcuts \cite{Pra86033405,Pra89053408},
(ii) multiple Schr\"{o}dinger pictures \cite{Prl109100403}, (iii) dressed-states-based shortcuts \cite{Prl116230503}, respectively.
This idea would be very promising in speeding up an adiabatic process without undesired couplings if one can
find out the suitable new paths for the system.
The main drawback of these approaches is the requirement of dynamical symmetry for the original Hamiltonian so far as the status.
It is still a problem to deal with a Hamiltonian that does not satisfy special dynamical symmetry.

Note that a set of suitable new paths always exist no matter the original Hamiltonian satisfies special dynamical symmetry or not.
The problem is how to find them out. In this paper, we will present a simple and straightforward way to
find the suitable new paths by constructing an easy-to-get intermediate Hamiltonian.
Now that the undesired off-diagonal terms are brought in by the supplementary Hamiltonian,
why not we create some imaginary terms in the original Hamiltonian to counteract the undesired off-diagonal terms?
Following this idea, we assume the intermediate Hamiltonian $\tilde{H}_{0}(t)$ is
formed by the original Hamiltonian $H_{0}(t)$ and the CD Hamiltonian $H_{cd}(t)$ with a
simple relationship, i.e., $\tilde{H}_{0}(t)=\bm{\lambda}(t)H_{0}(t)-\bm{\kappa}(t)H_{cd}(t)$, where
$\bm{\lambda}(t)$ and $\bm{\kappa}(t)$ are two operators which are used to correct the nonzero matrix elements
in $H_{0}(t)$ and $H_{cd}(t)$.
Then, by introducing suitable additional control fields,
it is possible to design an evolution exactly following the eigenvectors of the intermediate Hamiltonian.
The additional control fields will counteract undesired off-diagonal terms by appropriately choosing $\bm{\lambda}(t)$ and $\bm{\kappa}(t)$.
The eigenvectors of $\tilde{H}_{0}(t)$ should be ensured to coincide with the eigenvectors of $H_{0}(t)$
at the initial time ($t_{i}$) and the final time ($t_{f}$) in order to reproduce the same final state as an adiabatic process governed by $H_{0}(t)$, that is $\bm{\lambda}(t_{i})H_{0}(t_{i})\propto H_{0}(t_{i})$, $\bm{\lambda}(t_{f})H_{0}(t_{f})\propto H_{0}(t_{f})$, and
$\bm{\kappa}(t_{i})H_{cd}(t_{i})\simeq\bm{\kappa}(t_{f})H_{cd}(t_{f})\simeq0$.

In fact, the previous speedup schemes by using invariants \cite{Pra86033405,Pra89053408}, dressed states \cite{Prl116230503},
and multiple Schr\"{o}dinger pictures \cite{Prl109100403}
can be consider as recovered from the present approach with different choices of $\bm{\lambda}(t)$ and $\bm{\kappa}(t)$.
For example, when we suitably choose $\bm{\lambda}(t)$ and $\bm{\kappa}(t)$ so that
the final transitionless Hamiltonian $\tilde{H}(t)$ has the same
structure as the original Hamiltonian $H_{0}(t)$ and the eigenenergies of the intermediate Hamiltonian are time-independent,
the intermediate $\tilde{H}_{0}(t)$ can be regarded as an invariant of the system.
Which means the speedup schemes based on invariants
are actually special cases of the present approach.

\section{Review of the transitionless driving algorithm and the general problem}
The transitionless driving algorithm (TDA) shows that by adding a specific supplementary Hamiltonian
to the original Hamiltonian, the system's dynamics governed by the completed Hamiltonian will behave
ideally adiabatically along the eigenvectors of the original Hamiltonian.
Considering an arbitrary time-dependent Hamiltonian $H_{0}(t)$ satisfying
$H_{0}(t)|\phi_{n}(t)\rangle=E_{n}(t)|\phi_{n}(t)\rangle$,
where $|\phi_{n}(t)\rangle$ and $E_{n}(t)$ are the instantaneous eigenvectors and eigenenergies, respectively.
Under the adiabatic approximation, the state of the system at the time $t$ can be written as
$|\psi(t)\rangle=e^{i\beta_{n}(t)}|\phi_{n}(t)\rangle$,
where the adiabatic phase, with dynamical and geometric parts, is (the overdot means time derivative)
\begin{eqnarray}\label{eq0-33}
  \beta_{n}(t)=-\frac{1}{\hbar}\int_{0}^{t}dt'E_{n}(t')+i\int_{0}^{t}dt'\langle\phi_{n}(t')|\dot{\phi}_{n}(t')\rangle,
\end{eqnarray}
Then, defining a time-dependent unitary evolution $U=\sum_{n}e^{i\beta_{n}}|\phi_{n}(t)\rangle\langle\phi_{n}(0)|$
which must obey $i\hbar\dot{U}=H(t)U$, the transitionless Hamiltonian $H(t)$ to drive the system exactly along the adiabatic paths of $H_{0}(t)$ will be constructed as
$H(t)=H_{0}(t)+H_{cd}(t)$, where
\begin{eqnarray}\label{eq0-44}
  H_{cd}(t)=i\hbar\sum_{n}(|\dot{\phi}_{n}\rangle\langle\phi_{n}|-\langle\phi_{n}|\dot{\phi}_{n}\rangle|\phi_{n}\rangle\langle\phi_{n}|).
\end{eqnarray}
Here all kets are time-dependent. However, limited by the experimental technique, it has been pointed out that such
a transitionless Hamiltonian is usually hard to implement because of some undesired terms
introduced by $i\hbar\sum_{n}|\dot{\phi}_{n}\rangle\langle\phi_{n}|$.
The second term in $H_{cd}(t)$ only affects the phase and is in fact irrelevant to the following discussion, so
we ignore this term and assume $H_{cd}(t)=i\hbar\sum_{n}|\dot{\phi}_{n}\rangle\langle\phi_{n}|$.

Taking a three-level $\Lambda$-type system with an excited state $|2\rangle=[0,1,0]^{t}$
and two ground states $|1\rangle=[1,0,0]^{t}$ and $|3\rangle=[0,0,1]^{t}$ (superscript $t$ denotes the transpose) as an example,
the Hamiltonian describing a STIRAP
with resonant pump pulse $\Omega_{p}$ and Stokes pulse $\Omega_{s}$,
under the rotating wave approximation, is
$H_{0}(t)=\hbar\Omega_{p}|2\rangle\langle1|+\hbar\Omega_{s}|2\rangle\langle3|+H.c.$.
The corresponding instantaneous eigenvectors, with eigenenergies
$E_{0}=0$ and $E_{\pm}=\pm\hbar\Omega_{0}$ ($\Omega_{0}=\sqrt{\Omega_{p}^{2}+\Omega_{s}^{2}}$), are
$|\phi_{0}(t)\rangle=\cos{\theta}|1\rangle-\sin{\theta}|3\rangle$ and
$|\phi_{\pm}(t)\rangle=\frac{1}{\sqrt{2}}(\sin{\theta}|1\rangle\pm|2\rangle+\cos{\theta}|3\rangle)$, respectively,
where $\theta=\arctan{(\Omega_{p}/\Omega_{s})}$.
Under the adiabatic condition $|\dot{\theta}|\ll|\Omega_{0}|$,
perfect population transfer between the ground states $|1\rangle$ and $|3\rangle$
can be achieved by adiabatically following $|\phi_{0}(t)\rangle$ with the counterintuitive pulse order.
To speed up such a process, according to TDA \cite{Jpa42365303},
an auxiliary driving, $H_{cd}(t)=i\hbar\dot{\theta}|1\rangle\langle3|+H.c.$,
should be added to the original Hamiltonian $H_{0}(t)$.
That is, a resonant field with a specified phase connecting states $|1\rangle$ and $|3\rangle$ should be applied.
While, realizing such a 1-3 transition without causing other problems is still a challenge.
Firstly, the transition between $|1\rangle$ and $|3\rangle$ is usually electric-dipole forbidden.
Secondly, in some systems, i.e., cold atom systems, though one can use a microwave pulse to drive the transition between states $|1\rangle$ and $|3\rangle$,
the scheme is still hard to realize because it is very sensitive to the phase error of the 1-3 pulse \cite{Prl105123003,Natc712479}.

\section{Designing the intermediate Hamiltonian}
Here, we would like to note that by adding the CD Hamiltonian, the dimension $N$ of the system
has not been changed, so we can actually use the eigenvectors $\{|{\phi}_{n}(t)\rangle\}$ of the original Hamiltonian $H_{0}(t)$
to express another set of orthogonal
complete basis $\{|\tilde{\phi}_{n}(t)\rangle\}$ as $|\tilde{\phi}_{n}(t)\rangle=\sum_{m}C_{n,m}(t)|{\phi}_{m}(t)\rangle$,
where $C_{n,m}(t)$ is a time-dependent coefficient satisfying $\sum_{k}C_{n,k}^{*}(t)C_{m,k}(t)=\delta_{nm}$.
The time-dependent coefficients $\{C_{n,m}(t)\}$ can be used to form a unitary matrix $\mathbf{C}$,
in which the matrix element in the $n$th row and $m$th column is $C_{n,m}$.

Consider a special case that $\{|\tilde{\phi}_{n}(t)\rangle\}$ are the eigenvectors of Hamiltonian $\tilde{H}_{0}(t)$ with eigenenergies $\tilde{E}_{n}(t)$, respectively,
we might as well write $\tilde{H}_{0}(t)$ as $\tilde{H}_{0}(t)=\sum_{n}\tilde{E}_{n}(t)|\tilde{\phi}_{n}(t)\rangle\langle\tilde{\phi}_{n}(t)|$.
In this way, if the unitary matrix $\mathbf{C}$ is so special that $H_{0}(t)$ and $\tilde{H}_{0}(t)$ satisfies
\begin{align}\label{eq0-1}
  H_{0}(t)&=\tilde{H}_{0}(t)+\tilde{H}_{cd}(t), \cr
  \tilde{H}_{cd}(t)&=i\hbar\sum_{n}|\dot{\tilde{\phi}}_{n}(t)\rangle\langle\tilde{\phi}_{n}(t)|,
\end{align}
leading to
\begin{small}
\begin{align}\label{eq0-2}
  \sum_{n}{E}_{n}|{\phi}_{n}\rangle\langle{\phi}_{n}|=&
  \sum_{n,m,k}\tilde{E}_{n}C_{n,m}|{\phi}_{m}\rangle\langle{\phi}_{k}|C_{n,k}^{*} \cr
  &+i\hbar(\dot{C}_{n,m}|{\phi}_{m}\rangle+C_{n,m}|\dot{{\phi}}_{m}\rangle)C_{n,k}^{*}\langle{\phi}_{k}|.
\end{align}
\end{small}
The parameters should satisfy
\begin{align}\label{eq0-3}
  {E}_{l}\delta_{kl}=&\sum_{n}(\tilde{E}_{n}C_{n,k}C_{n,l}^{*}+i\hbar\dot{C}_{n,k}C_{n,l}^{*}\cr
                     &+i\hbar\sum_{m}C_{n,m}C_{n,l}^{*}\langle{\phi}_{k}|\dot{{\phi}}_{m}\rangle).
\end{align}

In fact, the idea of treating $\{|\tilde{\phi}_{n}(t)\rangle\}$ as the eigenvectors of $\tilde{H}_{0}(t)$ is highly consistent
with the previous approaches \cite{Prl109100403,Pra86033405,Prl116230503,Pra93052324}.
For example, $\{|\tilde{\phi}_{n}(t)\rangle\}$ can be understood as the dressed states mentioned in Ref. \cite{Prl116230503}.
Nevertheless, directly solving Eq. (\ref{eq0-3}) is really an outstanding challenge.
One can find that to satisfy the condition given by Eq. (\ref{eq0-3}), the unitary matrix $\mathbf{C}$ usually
has a rigid relationship with the original Hamiltonian $H_{0}(t)$ \cite{Prl109100403,Pra86033405,Prl116230503,Pra93052324}.
So directly finding solutions for Eq. (\ref{eq0-2}) is really a challenge
and seeking other ways to overcome the problem caused by the CD Hamiltonian is necessary.
Here, we would like to reemphasize the purpose of finding new paths for the system to evolve along with is
to overcome the problem caused by the undesired couplings in the CD Hamiltonian.
To realize this purpose, there are in fact two ways:

(i) Under the precondition that the transitionless Hamiltonian $H_{0}(t)$ is given with some kind of dynamical symmetry,
we can directly solve Eq. (\ref{eq0-3}) with a specific unitary matrix $\mathbf{C}$ like the previous works \cite{Prl109100403,Pra86033405,Prl116230503,Pra93052324}.

(ii) We can use a relatively simple unitary matrix $\mathbf{C}$ and orthogonal complete basis $\{|{\phi}_{n}(t)\rangle\}$ to construct
a new transitionless Hamiltonian $\tilde{H}(t)$ in form of the right hand side of Eq. (\ref{eq0-2}). Then we impose the undesired couplings in $\tilde{H}(t)$
to be inoperative (zero) in order to deduce the related parameters. In other words,
we can only focus on how to eliminate the undesired couplings in the right hand side of Eq. (\ref{eq0-2}).
Choosing suitable parameters, $\tilde{H}(t)$ has the same structure with $H_{0}(t)$.

Following the second idea, we write down all the matrix elements in the right hand side of Eq. (\ref{eq0-2}) as
\begin{eqnarray}\label{eq0-5}
  F_{l,r}&=&\langle\mu_{l}|[\sum_{n,m,k}\tilde{E}_{n}C_{n,m}|{\phi}_{m}\rangle\langle{\phi}_{k}|C_{n,k}^{*}
  +i\hbar(\dot{C}_{n,m}|{\phi}_{m}\rangle \cr
  &&+C_{n,m}|\dot{{\phi}}_{m}\rangle)C_{n,k}^{*}\langle{\phi}_{k}|]|\mu_{r}\rangle,
\end{eqnarray}
where $l,r=1,2,\cdots,N$ and $\{|\mu_{l}\rangle\}$ are the time-independent bare states of the system.
Consider a general case that the undesired couplings exist when $l=\alpha$ and $r=\beta$, we have
\begin{align}\label{eq0-6}
  F_{\alpha,\beta}
                 =&\sum_{n,m,k}\tilde{E}_{n}C_{n,m}{\phi}_{m,\alpha}{\phi}_{k,\beta}^{*}C_{n,k}^{*}\cr
                    &+i\hbar(\dot{C}_{n,m}{\phi}_{m,\alpha}+C_{n,m}\dot{{\phi}}_{m,\alpha})C_{n,k}^{*}{\phi}_{k,\beta}^{*},
\end{align}
leading to
\begin{align}\label{eq0-7a}
  \text{Re}(F_{\alpha,\beta})=&\text{Re}(F_{\beta,\alpha})=0,\cr
  \text{Im}(F_{\alpha,\beta})=&-\text{Im}(F_{\beta,\alpha})=0,
\end{align}
where ${\phi}_{n,b}$ ($b=\alpha,\beta$) means the $b$th element in the $n$th eigenvector $|{\phi}_{n}(t)\rangle$.
Here we would like to discuss a special case that $C_{n,m}=\delta_{nm}$ corresponding to $|\tilde{\phi}_{n}(t)\rangle=|\phi_{n}(t)\rangle$.
Noting that the undesired coupling is brought in by the CD Hamiltonian, which means
$\sum_{n}{{E}_{n}{\phi}_{n,\alpha}{\phi}_{n,\beta}^{*}}=0$. In this case, we can find if we change $E_{n}(t)$ to $\tilde{E}_{n}(t)$,
the imaginary part of the first term in the right hand side of Eq. (\ref{eq0-6}), $\sum_{n}{\tilde{E}_{n}{\phi}_{n,\alpha}{\phi}_{n,\beta}^{*}}$, always vanish.
While, the imaginary part of the second term in the right hand side of Eq. (\ref{eq0-6}), $\sum_{n}{i\hbar\dot{\phi}_{n,\alpha}\phi_{n,\beta}^{*}}$, is usually nonzero.
That is, there is usually no solution for Eq. (\ref{eq0-7a}) in case of $C_{n,m}=\delta_{nm}$.
Therefore, in order to ensure Eq. (\ref{eq0-7a}) mathematically solvable, we assume the first term in
the right hand side of Eq. (\ref{eq0-6}) $\sum_{n,m,k}{\tilde{E}_{n}C_{n,m}{\phi}_{m,\alpha}{\phi}_{k,\beta}^{*}C_{n,k}^{*}}$
has a special relationship with $\sum_{n}{i\hbar\dot{\phi}_{n,\alpha}\phi_{n,\beta}^{*}}$ that
\begin{align}\label{eq0-7b}
  \sum_{n,m,k}{\tilde{E}_{n}C_{n,m}{\phi}_{m,\alpha}{\phi}_{k,\beta}^{*}C_{n,k}^{*}}=-\kappa_{\alpha,\beta}\sum_{n}{i\hbar\dot{\phi}_{n,\alpha}\phi_{n,\beta}^{*}},
\end{align}
or we can express this relationship as
\begin{align}\label{eq0-7d}
  \langle\mu_{\alpha}|\tilde{H}_{0}(t)|\mu_{\beta}\rangle=-i\hbar\kappa_{\alpha,\beta}\langle\mu_{\alpha}|{H}_{cd}(t)|\mu_{\beta}\rangle,
\end{align}
where $\kappa_{\alpha,\beta}$ are time-dependent functions should be suitably chosen according to Eq. (\ref{eq0-7a}).
In this way, the problem of the undesired matrix elements is resolved.
Now we should focus on the other matrix elements.
Under the premise that structure of the new transitionless Hamiltonian $\tilde{H}(t)$ is similar to the original Hamiltonian $H_{0}(t)$,
it would be much better if the matrix elements $F_{l,r}$ ($l,r\neq\alpha,\beta$) satisfy
\begin{align}\label{eq0-7c}
  \sum_{n,m,k}{\tilde{E}_{n}C_{n,m}{\phi}_{m,l}{\phi}_{k,r}^{*}C_{n,k}^{*}}
          =\lambda_{l,r}\sum_{n}E_{n}\phi_{n,l}\phi_{n,r},
\end{align}
which corresponds to
\begin{align}\label{eq0-7f}
  \langle\mu_{l}|\tilde{H}_{0}(t)|\mu_{r}\rangle=\lambda_{l,r}\langle\mu_{l}|{H}_{0}(t)|\mu_{r}\rangle,
\end{align}
where $\lambda_{l,r}$ are also time-dependent functions. Then,
in order to coincide $\{|\tilde{\phi}_{n}(t)\rangle\}$ with $\{{|\phi}_{n}(t)\rangle\}$ at the initial time ($t_{i}$) and
the final time ($t_{f}$), $\lambda_{l,r}$ and $\kappa_{\alpha,\beta}$ should be suitably chosen to satisfy
$\mathbf{C}(t_{i})=\mathbf{C}(t_{f})=I$ ($I$ is the identity matrix).
Thus, all of the matrix elements for the new transitionless Hamiltonian $\tilde{H}(t)$
have been suitably chosen, and for the sake of convenience,
we might write the Hamiltonian $\tilde{H}_{0}(t)$ as
\begin{eqnarray}\label{eq2a-1}
  \tilde{H}_{0}(t)=\bm{\lambda}(t)H_{0}(t)-\bm{\kappa}(t)H_{cd}(t).
\end{eqnarray}
according to Eqs. (\ref{eq0-7b}) and (\ref{eq0-7c}),
where $\bm{\lambda}(t)$ and $\bm{\kappa}(t)$ are two operators which are used to correct the nonzero matrix elements
in $H_{0}(t)$ and $H_{cd}(t)$.
Then, one can drive the system to evolve exactly along the eigenvector $|\tilde{\phi}_{n}(t)\rangle$
with Hamiltonian
\begin{align}\label{eq0-8}
  \tilde{H}(t)=&\tilde{H}_{0}(t)+\tilde{H}_{cd}(t), \cr
  \tilde{H}_{cd}(t)=&i\hbar\sum_{n}|\dot{\tilde{\phi}}_{n}(t)\rangle\langle\tilde{\phi}_{n}(t)|.
\end{align}

\section{Application example for a STIRAP Hamiltonian with off-resonant driving fields}
We apply our general approach to a three-level $\Lambda$-type system with an excited state $|2\rangle$ and
two ground states $|1\rangle$ and $|3\rangle$ which are off-resonantly driven by two classical pulses.
The original Hamiltonian within the rotating-wave approximation (RWA) reads
\begin{eqnarray}\label{eq3-1}
  H_{0}(t)=\frac{\hbar}{2}\left(
                \begin{array}{ccc}
                  0 & \Omega_{p}(t) & 0 \\
                  \Omega_{p}(t) & 2\Delta(t) & \Omega_{s}(t) \\
                  0 & \Omega_{s}(t) & 0
                \end{array}
           \right),
\end{eqnarray}
where $\Omega_{p}(t)$, $\Omega_{s}(t)$, and $\Delta(t)$ are the pump pulse, Stokes pulse, and detuning, respectively.
When the detuning is nonzero but is not large, there are still no effective ways
to construct shortcuts for this system without the undesired 1-3 pulse.
The previous works focus on the resonant or large detuning case \cite{Prl109100403,Pra86033405,Prl116230503,Natc712479}.
For resonant driving, the system evolution is subject to experimental imperfections.
Therefore, a slightly detuned driving configuration may be a better choice to realize the desired speedup dynamics in practice.

We will show in the following how we can use the present approach to construct shortcuts for this off-resonant system without
the undesired 1-3 pulse. The corresponding eigenvectors for the original Hamiltonian $H_{0}(t)$ are
$|\phi_{0}(t)\rangle=\cos{\theta}|1\rangle-\sin{\theta}|3\rangle$,
$|\phi_{+}(t)\rangle=\sin{\theta}\sin{\varphi}|1\rangle+\cos{\varphi}|2\rangle+\cos{\theta}\sin{\varphi}|3\rangle$,
and $|\phi_{-}(t)\rangle=\sin{\theta}\cos{\varphi}|1\rangle-\sin{\varphi}|2\rangle+\cos{\theta}\cos{\varphi}|3\rangle$,
with eigenenergies $E_{0}(t)=0$, $E_{+}(t)=\hbar\Xi_{0}\cos^{2}{\varphi}$, and $E_{-}(t)=-\hbar\Xi_{0}\sin^{2}{\varphi}$,
where $\Xi_{0}=\sqrt{\Omega_{0}^{2}+\Delta^{2}(t)}$.
The parameters $\theta$ and $\varphi$ are defined by $\tan{\theta}=\Omega_{p}(t)/\Omega_{s}(t)$ and
$\tan{2\varphi}=\Omega_{0}/\Delta(t)$, respectively.
The CD Hamiltonian for the present system is
\begin{eqnarray}\label{eq3-3}
  H_{cd}(t)=i\hbar\left(
                \begin{array}{ccc}
                  0 & \dot{\varphi}\sin{\theta} & \dot{\theta} \\
                  -\dot{\varphi}\sin{\theta} & 0 & -\dot{\varphi}\cos{\theta} \\
                  -\dot{\theta} & \dot{\varphi}\cos{\theta} & 0
                \end{array}
           \right).
\end{eqnarray}
Then, according to Eq. (\ref{eq2a-1}), we have
\begin{small}
\begin{align}\label{eq3-5}
  \bm{\lambda}(t)H_{0}(t)&=\frac{\hbar}{2}\left(
                \begin{array}{ccc}
                  0 & \lambda_{p}\Omega_{p}(t) & 0 \\
                  \lambda_{p}\Omega_{p}(t) & 2\lambda_{d}\Delta(t) & \lambda_{s}\Omega_{s}(t) \\
                  0 & \lambda_{s}\Omega_{s}(t) & 0
                \end{array}
           \right), \cr
  \bm{\kappa}(t)H_{cd}(t)&=i\hbar\left(
                \begin{array}{ccc}
                  0 & \kappa_{p}\dot{\varphi}\sin{\theta} & \kappa_{a}\dot{\theta} \\
                  -\kappa_{p}\dot{\varphi}\sin{\theta} & 0 & -\kappa_{s}\dot{\varphi}\cos{\theta} \\
                  -\kappa_{a}\dot{\theta} & \kappa_{s}\dot{\varphi}\cos{\theta} & 0
                \end{array}
           \right),
\end{align}
\end{small}
with $\lambda_{p,(s,d)}$ and $\kappa_{p,(s,a)}$ being time-dependent real coefficients.
To satisfy the condition $\bm{\lambda}(t_{i})\simeq\bm{\lambda}(t_{f})\simeq$const and
$\bm{\kappa}(t_{i})H_{cd}(t_{i})\simeq\bm{\kappa}(t_{f})H_{cd}(t_{f})\simeq0$, we may choose a simple set of $\lambda_{p,(s,d)}$ and $\kappa_{p,(s,a)}$ as
\begin{align}\label{eq3-6}
  \lambda_{p}\Xi_{0}&=\lambda_{s}\Xi_{0}={\tilde{\Xi}_{0}}\cos{\gamma},\cr
  \lambda_{d}\Xi_{0}&={\tilde{\Xi}_{0}\cos{2\gamma}}/({\cos^{2}{\gamma}}), \cr
  \kappa_{p}\dot{\varphi}&=\tilde{\Xi}_{0}\cot{\theta}\tan{\gamma}\cos{2\varphi},\cr
  \kappa_{s}\dot{\varphi}&=-\tilde{\Xi}_{0}\tan{\theta}\tan{\gamma}\cos{2\varphi}, \cr
  \kappa_{a}\dot{\theta}&=\tilde{\Xi}_{0}\sin{\gamma}\sin{2\varphi}/2,
\end{align}
where $\tilde{\Xi}_{0}$ is an energy parameter related to the
eigenenergies of the original Hamiltonian and $\gamma$
is to be chosen to satisfy $\gamma(t_{i})=\gamma(t_{f})=0$.
With this, the eigenvectors of the intermediate Hamiltonian are
\begin{eqnarray}\label{eq3-7}
  |\tilde{\phi}_{0}(t)\rangle&=&\left(
                                   \begin{array}{c}
                                     \cos{\theta}\cos{\gamma} \\
                                     -i\sin{\gamma} \\
                                     -\sin{\theta}\cos{\gamma}
                                   \end{array}
                              \right),
                              \cr\cr
  |\tilde{\phi}_{+}(t)\rangle&=&\left(
                                   \begin{array}{c}
                                     \sin{\theta}\sin{\varphi}-i\cos{\theta}\cos{\varphi}\sin{\gamma} \\
                                     \cos{\varphi}\cos{\gamma} \\
                                     \cos{\theta}\sin{\varphi}+i\sin{\theta}\cos{\varphi}\sin{\gamma}
                                   \end{array}
                              \right),
                              \cr\cr
  |\tilde{\phi}_{-}(t)\rangle&=&\left(
                                   \begin{array}{c}
                                     \sin{\theta}\cos{\varphi}+i\cos{\theta}\sin{\varphi}\sin{\gamma} \\
                                     -\sin{\varphi}\cos{\gamma} \\
                                     \cos{\theta}\cos{\varphi}-i\sin{\theta}\sin{\varphi}\sin{\gamma}
                                   \end{array}
                              \right),
\end{eqnarray}
with the eigenenergies $\tilde{E}_{0}(t)=\hbar\tilde{\Xi}_{0}(\sin^{4}{\varphi}-\cos^{4}{\varphi})\tan^{2}{\gamma}$,
$\tilde{E}_{+}(t)=\hbar\tilde{\Xi}_{0}\cos^{2}{\varphi}$, and $\tilde{E}_{-}(t)=-\hbar\tilde{\Xi}_{0}\sin^{2}{\varphi}$,  respectively.
Then, we have ($\alpha=1,\ \beta=3$)
\begin{eqnarray}\label{eq3-8}
  F_{1,3}=\dot{\theta}+i\dot{\varphi}\sin{\gamma}\cos{2\theta}-\tilde{\Xi}_{0}\sin{\gamma}\sin{2\varphi}/2.
\end{eqnarray}
Obviously, when $\dot{\varphi}=0$ and $\tilde{\Xi}_{0}=2\dot{\theta}/(\sin{\gamma}\sin{2\varphi})$,
the condition $F_{1,3}=0$ is satisfied and the undesired 1-3 pulse is no longer required.
The final transitionless Hamiltonian reads
\begin{small}
\begin{eqnarray}\label{eq3-9}
  \tilde{H}(t)=\frac{\hbar}{2}\left(
                \begin{array}{ccc}
                  0 & \tilde{\Omega}_{p}(t)e^{i\vartheta_{p}(t)} & 0 \\
                  \tilde{\Omega}_{p}(t)e^{-i\vartheta_{p}(t)} & 2\tilde{\Delta}(t) & \tilde{\Omega}_{s}(t)e^{-i\vartheta_{s}(t)} \\
                  0 & \tilde{\Omega}_{s}(t)e^{i\vartheta_{s}(t)} & 0
                \end{array}
           \right),
\end{eqnarray}
\end{small}
where the Rabi frequencies $\tilde{\Omega}_{p,(s)}(t)$, phase shifts $\vartheta_{p,(s)}(t)$
and detuning $\tilde{\Delta}(t)$ for the final transitionless Hamiltonian are
\begin{eqnarray}\label{eq3-10}
  \tilde{\Omega}_{p}(t)&=&[(\tilde{\Xi}_{0}\cos{\gamma}\sin{\theta}\sin{2\varphi}+\dot{\gamma}\cos{\theta})^{2}\cr
                         &&+(2\tilde{\Xi}_{0}\tan{\gamma}\cos{\theta}\cos{2\varphi})^{2}]^{1/2},\cr\cr
  \tilde{\Omega}_{s}(t)&=&[(\tilde{\Xi}_{0}\cos{\gamma}\cos{\theta}\sin{2\varphi}-\dot{\gamma}\sin{\theta})^{2}\cr
                         &&+(2\tilde{\Xi}_{0}\tan{\gamma}\sin{\theta}\cos{2\varphi})^{2}]^{1/2},\cr\cr
  \vartheta_{p}(t)&=&\arctan{\left(\frac{2\tilde{\Xi}_{0}\tan{\gamma}\cos{\theta}\cos{2\varphi}}
                      {\tilde{\Xi}_{0}\cos{\gamma}\sin{\theta}\sin{2\varphi}+\dot{\gamma}\cos{\theta}}\right)},\cr\cr
  \vartheta_{s}(t)&=&\arctan{\left(\frac{2\tilde{\Xi}_{0}\tan{\gamma}\sin{\theta}\cos{2\varphi}}
                      {\tilde{\Xi}_{0}\cos{\gamma}\cos{\theta}\sin{2\varphi}-\dot{\gamma}\sin{\theta}}\right)},\cr\cr
  \tilde{\Delta}(t)&=&\tilde{\Xi}_{0}\cos{2\gamma}\cos{2\varphi}/\cos^{2}{\gamma},
\end{eqnarray}
respectively. Here we focus on the discussion for the Rabi frequencies $\tilde{\Omega}_{p}(t)$ and $\tilde{\Omega}_{s}(t)$.
We can in fact write $\tilde{\Omega}_{p,(s)}(t)$ as
\begin{align}\label{eqS16}
  \tilde{\Omega}_{p}(t)=\tilde{\Omega}_{0}\sin{\tilde{\theta}}, \ \tilde{\Omega}_{s}(t)=\tilde{\Omega}_{0}\cos{\tilde{\theta}},
\end{align}
where the modified pulses' amplitude and angle are
\begin{widetext}
\begin{align}\label{eqS17}
  \tilde\Omega_{0}&=\sqrt{(\tilde{\Xi}_{0}\cos{\gamma}\sin{2\varphi})^{2}+(2\tilde{\Xi}_{0}\tan{\gamma}\cos{2\varphi})^{2}+\dot{\gamma}^{2}},\cr
  \tilde{\theta}&=\arctan\left[\sqrt{\frac{(\tilde{\Xi}_{0}\cos{\gamma}\sin{\theta}\sin{2\varphi}+\dot{\gamma}\cos{\theta})^{2}
                         +(2\tilde{\Xi}_{0}\tan{\gamma}\cos{\theta}\cos{2\varphi})^{2}}{(\tilde{\Xi}_{0}\cos{\gamma}\cos{\theta}\sin{2\varphi}-\dot{\gamma}\sin{\theta})^{2}
                         +(2\tilde{\Xi}_{0}\tan{\gamma}\sin{\theta}\cos{2\varphi})^{2}}}\right].
\end{align}
We consider a special case that the detuning $\Delta(t)$ is so small that $\varphi\Rightarrow\pi/4$. Then, we have
\begin{align}\label{eqS18}
  \tilde\Omega_{0}\approx\sqrt{(\tilde{\Xi}_{0}\cos{\gamma})^{2}+\dot{\gamma}^{2}},\
  \tilde{\theta}\approx\theta+\arctan[\dot{\gamma}/(\tilde{\Xi}_{0}\cos{\gamma})].
\end{align}
Then according to $\tilde{\Xi}_{0}=2\dot{\theta}/(\sin{\gamma}\sin{2\varphi})$,
we can set
\begin{align}\label{eqS19}
  \gamma=\arctan{[\dot{\theta}/\sqrt{(\lambda_{p}\Omega_{p})^{2}+(\lambda_{s}\Omega_{s})^{2}}]}, \
  \tilde{\Xi}_{0}\cos{\gamma}\approx 2\sqrt{(\lambda_{p}\Omega_{p})^{2}+(\lambda_{s}\Omega_{s})^{2}}.
\end{align}
\end{widetext}
We find, Eq. (\ref{eqS17}) can be further written as
\begin{align}\label{eqS20}
  \tilde\Omega_{0}&\approx\sqrt{4\lambda_{p}^{2}\Omega_{p}^{2}+4\lambda_{s}^{2}\Omega_{s}^{2}+\dot{\gamma}^{2}},\cr
  \tilde{\theta}&\approx\theta+\arctan[\dot{\gamma}/(2\sqrt{\lambda_{p}^{2}\Omega_{p}^{2}+\lambda_{s}^{2}\Omega_{s}^{2}})].
\end{align}
This is just the result of the previous approach in Ref. \cite{Prl116230503}.

\begin{figure}
 \scalebox{0.3}{\includegraphics {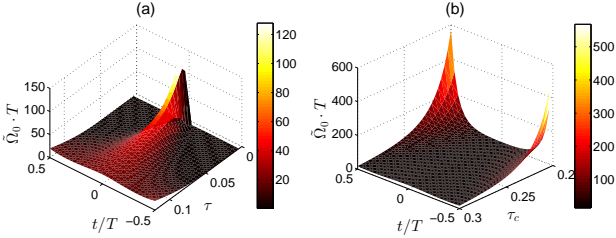}}
 \caption{
         (a) The relationship between the dimensionless parameter $T\tilde{\Omega}_{0}$ and $\tau$.
         The maximum value for $T\tilde{\Omega}_{0}$ is when $t=0$, and $T\tilde{\Omega}_{0}^{max}$
         observably decreases with the increasing of $\tau$  as shown in the figure.
         (b) The relationship between the dimensionless parameter $T\tilde{\Omega}_{0}$ and $\tau_{c}$.
         The maximum value for $T\tilde{\Omega}_{0}$ is when $t=\pm T/2$, and $T\tilde{\Omega}_{0}^{max}$
         also decreases with the increasing of $\tau_{c}$ as shown in the figure. We choose $\tau_{c}=0.3T$ and $\tau=0.12T$ in plotting
         Figs. 1 (a) and (b), respectively. Other parameters are $\gamma_{0}=0.1$ and $\varphi=\pi/5$.
         }
 \label{Figa1}
\end{figure}

In order to design a reference process to obtain the target state
$|3\rangle$ from the initial state $|1\rangle$ through the evolution path $|\tilde{\phi}_{0}(t)\rangle$,
we can accordingly set the boundary condition as $\theta(t_{i})=0,\ \theta(t_{f})=\pi/2$ and
$\gamma(t_{i})=\gamma(t_{f})=0$.
Therefore, we consider the optimal STIRAP pulses discussed in Ref. \cite{Pra80013417},
\begin{align}\label{eqS21}
  \Omega_{0}=\chi e^{-(t/T_{0})^{2n}}, \ \theta=\frac{\pi}{2(1+e^{-t/\tau})},
\end{align}
with $\chi$ being the peak of the pulses. To satisfy the boundaries $\theta(t_{i})=0$ and $\theta(t_{f})=\pi/2$, $\tau$ should satisfies $0<\tau\leq 0.12T$
($T=t_{f}-t_{i}$ is the total operation time and $\theta(t_{i})\approx0.07\pi$ when $\tau=0.12T$).
For the angle $\gamma$, the boundaries are $\gamma(t_{i})=\gamma(t_{f})=0$ and $\dot{\gamma}(t_{i})=\dot{\gamma}(t_{f})=0$
to simulate the modified pulses with a finite duration.
The simplest choice for $\gamma$ can be Gaussian function that
\begin{align}\label{eqS22}
  \gamma=\pi\gamma_{0}\cdot\exp(-t^{2}/\tau_{c}^{2}),
\end{align}
with $0<\gamma_{0}<0.5$ and $0.2T<\tau_{c}<0.3T$.
Then we have
\begin{align}\label{eqS23}
  \dot{\theta}=\frac{\pi}{2}\cdot\frac{e^{-t/\tau}}{\tau(1+e^{-t/\tau})^{2}},\ \dot{\gamma}=-\frac{2\pi\gamma_{0}t}{\tau_{c}^{2}}e^{-t^{2}/\tau_{c}^2}.
\end{align}

\begin{figure}
 \scalebox{0.33}{\includegraphics {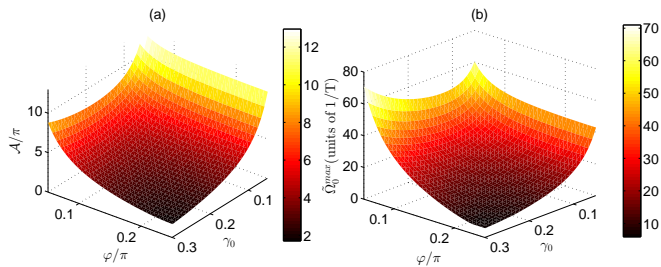}}
 \caption{
         (a) The total pulse area $\mathcal{A}$ given by Eqs. (\ref{eqS16}) and (\ref{eqS25})
         versus the parameters $\gamma_{0}$ and $\varphi$.
         (b) $T\tilde{\Omega}_{0}^{max}$ versus the parameters $\gamma_{0}$ and $\varphi$, where
         $\tilde{\Omega}_{0}^{max}$ means the maximal value for the modified pulse amplitude $\tilde{\Omega}_{0}$.
         Other parameters are $\tau=0.115T$ and $\tau_{c}=0.3T$.
         }
 \label{figS2}
\end{figure}

\begin{figure}
 \scalebox{0.3}{\includegraphics {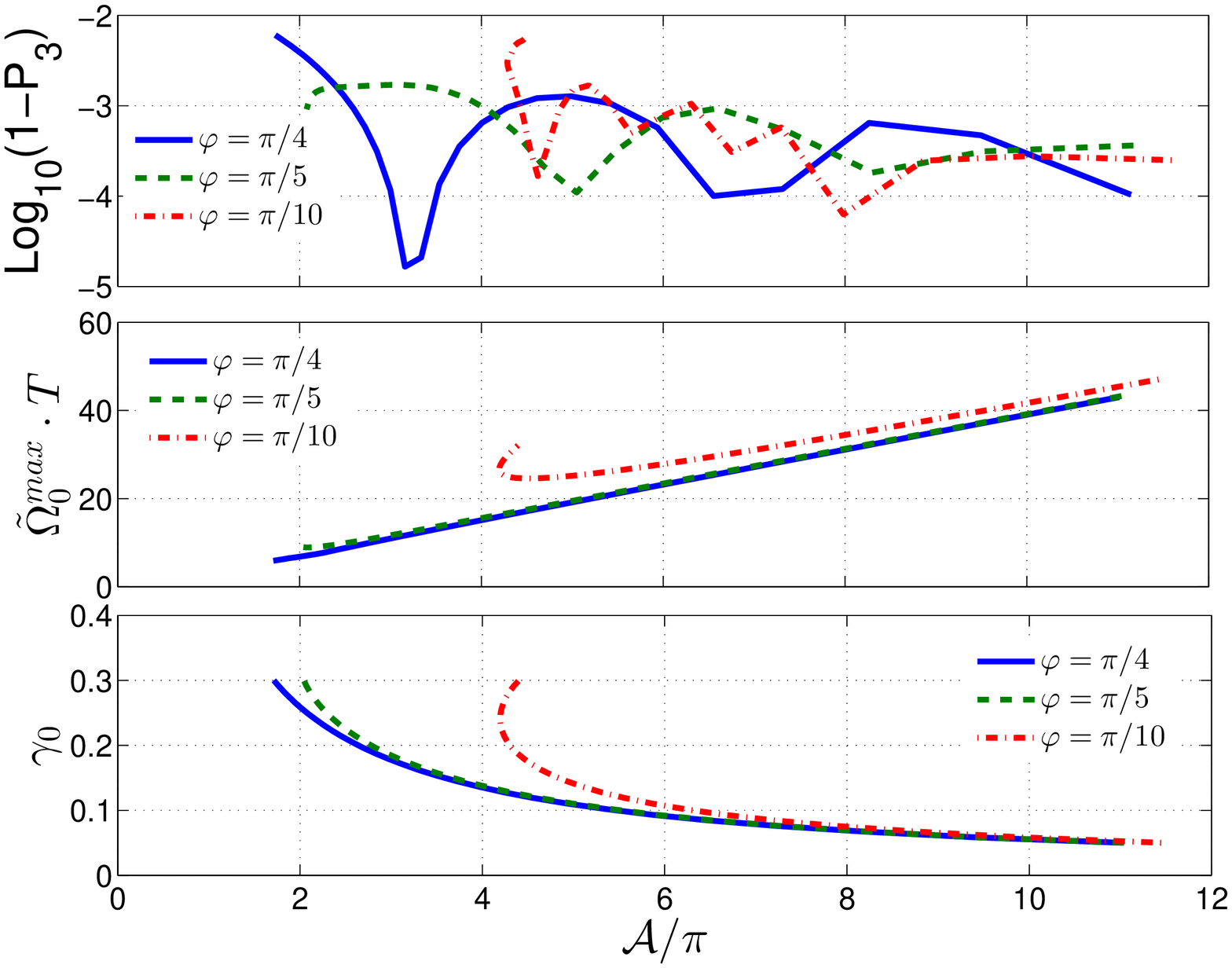}}
 \caption{
         Upper frame: Logarithmic scale of the deviation from the complete population transfer with different choices for the
         parameter $\varphi$.
         Middle frame: The corresponding dimensionless parameter $T\tilde{\Omega}_{0}^{max}$ for the speedup process.
         Lower frame: The relationship between the parameter $\gamma_{0}$
         (corresponding to the maximal population for the excited state $|2\rangle$) and the total pulse area $\mathcal{A}$.
         Other parameters are $\tau=0.115T$, $\tau_{c}=0.3T$ according to the requirement of less energy cost.
         }
 \label{figS3}
\end{figure}

To analyze total interaction time for a scheme, in general, the
dimensionless parameter $T\tilde{\Omega}_{0}^{max}$ ($\tilde{\Omega}_{0}^{max}$
is the maximum value of $\tilde{\Omega}_{0}$) is a common measurement scale.
Beware that $\tilde{\Omega}_{0}^{max}$ is usually a little larger than the maximum value for $\tilde{\Omega}_{p,(s)}(t)$,
the total interaction time measured by the $T\tilde{\Omega}_{0}^{max}$ is in fact a little larger than the real one.
While, $T\tilde{\Omega}_{0}^{max}$ would help a lot for quantitative analysis in the total interaction time, so, we tend to use $T\tilde{\Omega}_{0}^{max}$
as a measurement scale for the total interaction time in the following discussion.
Then, with ${\theta}$ and ${\gamma}$ given in Eqs. (\ref{eqS21}) and (\ref{eqS22}),
we find, within the selectable range for the parameters, $T\tilde{\Omega}_{0}^{max}$ is in inverse proportion to $\tau$ and $\tau_{c}$.
Numerical simulation as shown in Fig. \ref{Figa1} which displays the relationship between
$T\tilde{\Omega}_{0}$ and $\tau$ [see Fig. \ref{Figa1} (a)], the relationship between
$T\tilde{\Omega}_{0}$ and $\tau_{c}$ [see Fig. \ref{Figa1} (b)], also verifies this point.
That is, $\tau$ and $\tau_{c}$ should be chosen the larger the better to shorten the interaction time.
Therefore, we might choose the largest $\tau=0.115T$ and $\tau_{c}=0.3T$ for the preliminary discussion of the approach.
To analyze the energy cost of the process, we define the total pulse area for the process as
\begin{align}\label{eqS24}
  \mathcal{A}=\int_{-\infty}^{+\infty}dt\sqrt{\tilde{\Omega}_{p}^{2}(t)+\tilde{\Omega}_{s}^{2}(t)}.
\end{align}
For the pulses in Eq. (\ref{eqS16}), the total pulse area is
\begin{align}\label{eqS25}
  \mathcal{A}=\int_{-\infty}^{+\infty}dt\tilde{\Omega}_{0}(t).
\end{align}
Fig. \ref{figS2} (a) shows the total pulse area $\mathcal{A}$ versus $\gamma_{0}$ and $\varphi$.
As we can find, more energy should be cost with both the decreasing of $\varphi$ (corresponding to the increasing of the detuning) and
the decreasing of $\gamma_{0}$ (corresponding to the decreasing of the excited-state population) as shown in the figure.
In Fig. \ref{figS2} (b), we plot the dimensionless parameter $T\tilde{\Omega}_{0}^{max}$ versus $\gamma_{0}$ and $\varphi$.
Obviously, the total interaction time increases with the increasing of detuning as shown in Fig. \ref{figS2} (b).
The deviation from the complete population
transfer displayed in Fig. \ref{figS3} (the upper frame) with a logarithmic scale
shows the high efficiency for the present approach.
We obtain the transfer with an accuracy to about four digits for
areas are: (i) $2.9\pi$ when $\varphi=\pi/4$ (the resonant case), (ii) $4.9\pi$ when $\varphi=\pi/5$ (the small detuning case),
(iii) $7.9\pi$ when $\varphi=10$ (large detuning case), respectively. The corresponding total
interaction time as shown in the middle frame in Fig. \ref{figS3}
are: (i) $T\tilde{\Omega}_{0}^{max}=11.5$ when $\mathcal{A}=2.9\pi$ for the resonant case,
(ii) $T\tilde{\Omega}_{0}^{max}=19.5$ when $\mathcal{A}=4.9\pi$ for the small detuning case,
(iii) $T\tilde{\Omega}_{0}^{max}=34$ when $\mathcal{A}=7.9\pi$ for the large detuning case.
It is known that in the $\Lambda$-type system, the use of two
successive $\pi$ pulses, respectively, for the pump and Stokes
fields, or of two overlapping fields corresponds both to $\mathcal{A}=2\pi$ \cite{Aps58243}, and the minimum area in a $\Lambda$-type system for
a population transfer from $|1\rangle$ to $|3\rangle$ is $\sqrt{3}\pi$ \cite{Jmp432107}.
The total pulse area $\mathcal{A}$ in the present approach as shown in Figs. \ref{figS2} and \ref{figS3},
is small enough to verify the state transfer is sped up.
Then, in the lower frame in Fig. \ref{figS3}, we give the relationship between the parameter $\gamma_{0}$
and $\mathcal{A}$. Clearly from the figure, $\gamma_{0}$ generally decreases with the
increasing of $\mathcal{A}$, which means, an increasing in the population of intermediate state
is unavoidable if one wants to save energy to drive the system to the target state.
This result can be understood by, the intermediate state links the whole system together just like brittle
strings; the evolution of the system is interdictory without
the participation of these intermediate states. By increasing the
populations of intermediate states in a certain period of time,
just like broadening the path for the transition between
the initial state and the target state in a certain period of time, the evolution could
be much faster.

\begin{figure}
 \scalebox{0.32}{\includegraphics {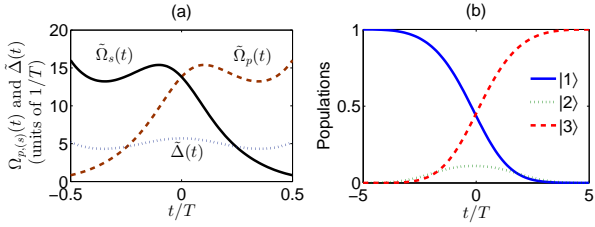}}
 \caption{
         (a) The modified Rabi frequencies and detuning [Eq. (\ref{eq3-1})] versus time.
         (b) The speedup population transfer governed by the off-resonant $\Lambda$-type system with Hamiltonian $\tilde{H}(t)$.
         Parameters are $\tau_{c}=0.3T$, $\tau=0.115T$, $\gamma_{0}=0.1$, and $\varphi=\pi/5$. With
         these parameters, the maximal pulse amplitude for $\tilde{H}(t)$ is $\tilde{\Omega}_{0}^{max}\approx16/T$.
         }
 \label{fig1}
\end{figure}

\begin{figure}
 \scalebox{0.32}{\includegraphics {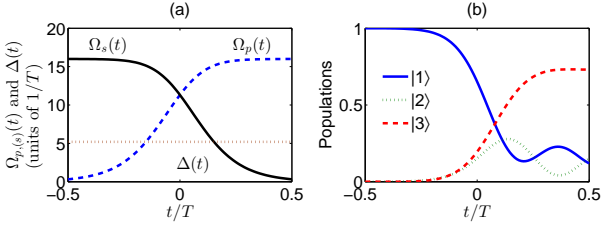}}
 \caption{
         (a) The original Rabi frequencies and detuning [Eq. (\ref{eq3-10})] versus time.
         (b) The population transfer governed by the off-resonant $\Lambda$-type system with the original Hamiltonian ${H_{0}}(t)$.
         Parameters are $\tau_{c}=0.3T$, $\tau=0.115T$, and $\varphi=\pi/5$.
         The modified pulse amplitude $\tilde{\Omega}_{0}$ is in fact independent to the original pulse amplitude $\Omega_{0}$ according to
         Eq. (\ref{eq3-1}). So, for a fair comparison, we choose a relatively larger $\Omega_{0}=16/T$ in plotting Fig. 4.
         }
 \label{fig2}
\end{figure}

Then, with parameters $\{\gamma_{0}=0.1,\tau=0.115T,\tau_{c}=0.3T,\varphi=\pi/5\}$
(choosing $\gamma_{0}=0.1$ for a less excited-state population), we show the modified pulses $\tilde{\Omega}_{p,(s)}(t)$
and detuning $\tilde{\Delta}(t)$ in Fig. \ref{fig1} (a). The maximal
amplitude for the pulses in this case $\tilde{\Omega}_{0}^{max}\approx 16/T$ with pulse area $\mathcal{A}\approx4.1\pi$.
The dynamics governed by the final transitionless
Hamiltonian $\tilde{H}(t)$ (with $\tau\approx0.115T$) versus time is shown in Fig. \ref{fig1} (b).
For comparison, with the same parameters for the pulses and detuning,
we accordingly plot the original parameters for the original Hamiltonian $H_{0}(t)$ and the time evolution of the system
in Figs. \ref{fig2} (a) and (b), respectively. We choose $\Omega_{0}=16/T$ in plotting Fig. \ref{fig2}. The total pulse area $\mathcal{A}$
corresponding to Fig. \ref{fig2} is $\mathcal{A}=\int_{-\infty}^{+\infty}dt\Omega_{0}=5.09\pi$.
The comparison between Fig. \ref{fig1} (b) and Fig. \ref{fig2} (b) shows, within a specified
total evolution time $T$, the final transitionless
Hamiltonian $\tilde{H}(t)$ allows a near-perfect population transfer (with $P_{3}\approx0.997$) from $|1\rangle$ to $|3\rangle$ while the
original Hamiltonian $H_{0}(t)$ fails. In fact, to satisfy the adiabatic condition
$\dot{\theta}\ll|\Xi_{0}\sin^{2}{\varphi}/\cos{\varphi}|<|\Xi_{0}\cos^{2}{\varphi}/\sin{\varphi}|$ (assuming $0<\varphi<\pi/4$)
for the original Hamiltonian $H_{0}(t)$, with the $\theta$ given in Eq. (\ref{eqS21}) and $\varphi=\pi/5$,
$\Omega_{0}$ should be at least $30/T$ to ensure $(\Xi_{0}\sin^{2}{\varphi})/(\dot{\theta}\cos{\varphi})\geq 5$.

\begin{figure}
 \scalebox{0.32}{\includegraphics {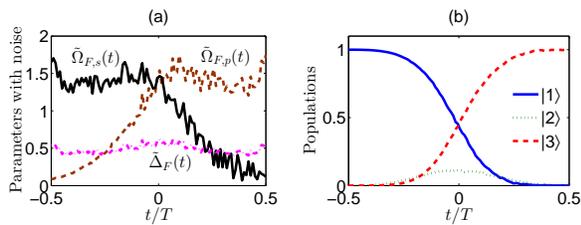}}
 \caption{
         (a) The shapes for the modified pulses and detuning when instantaneous fluctuations are taken into consideration according to Eq. (\ref{eq3-12a}).
         (b) The population transfer with the fluctuation-included parameters. Shown in the figure, even when
         the shapes for the modified pulses and detuning are destroyed seriously, a nearly perfect population transfer (with $P_{3}\approx 0.993$)
         still can be achieved. Which verifies that the scheme is robust against the instantaneous fluctuations on the parameters.
         }
 \label{fig3}
\end{figure}

\begin{figure}
 \scalebox{0.23}{\includegraphics {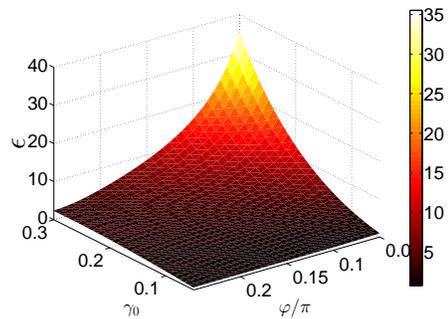}}
 \caption{
         The measurement scale $\epsilon$ defined by $\epsilon=\Gamma_{a}\cdot\overline{P}_{2}\cdot(\tilde{\Omega}_{0}^{max}\cdot T)$
         for analyzing the influence of spontaneous emission versus $\gamma_{0}$ and $\varphi$. Parameters are chosen as $\tau=0.115T$, $\tau_{c}=0.3T$,
         and $\Gamma_{a}=0.5$ in plotting the figure. For a robust process, $\epsilon$ is the smaller the better, so a very large $\gamma_{0}$ which
         though significantly shortens the interaction time, and a very small $\varphi$ which though restrains the excited-state population,
         are certainly both not desirable.
         }
 \label{figS4}
\end{figure}

\begin{figure}
 \scalebox{0.23}{\includegraphics {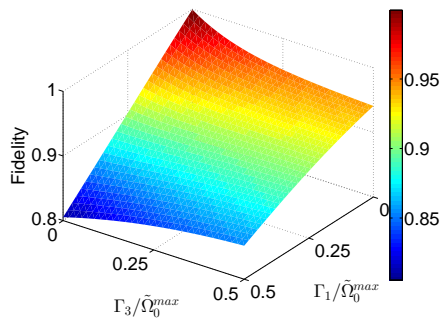}}
 \caption{
         The fidelity defined by $|\langle 3|\rho(t_{f})|3\rangle|^{2}$ for the population transfer versus $\Gamma_{1}$ and $\Gamma_{3}$
         in an off-resonant system with small detuning. Parameters are chosen as $\tau=0.115T$, $\tau_{c}=0.3T$, $\gamma_{0}=0.1$, and $\varphi=\pi/5$.
         In this case, the interaction time $T\approx 16/\tilde{\Omega}_{0}^{max}$, the total pulse area $\mathcal{A}\approx 4.1\pi$.
         }
 \label{fig4}
\end{figure}

We know that the Vitanov pulses were originally derived in order to get an ultra-high and robust fidelity
in an adiabatic process \cite{Pra80013417,Pra80043408}. To check the robustness against
noise after the correction to the pulses, we define a parameter with noise as
\begin{align}\label{eq3-12a}
  {G}_{F}=G+G\cdot\varpi,
\end{align}
where $G=\tilde{\Omega}_{0},\ \tilde{\theta},\ \tilde{\Delta}(t)$, and $-0.1<\varpi<0.1$ is a random number.
This allows one to model instantaneous fluctuations of the detunings and of the pulse shapes.
We can find from Fig. \ref{fig3} (a) that when such a noise is considered, the shapes for
$\tilde{\Omega}_{p,(s)}$ and $\tilde{\Delta}(t)$ have been destroyed seriously. However, a nearly perfect
population transfer (with the final population for $|3\rangle$ is $P_{3}(t_{f})\approx0.993$) still can be achieved as shown in Fig. \ref{fig3} (b). Which means
the present approach with modified Vitanov pulses is still robust against instantaneous
fluctuations on the parameters. The main decoherence effect in the population transfer is due to the
spontaneous emission from the excited state $|2\rangle$.
The total spontaneous emission from the excited state, is generally decided by two
factors: the total interaction time and the average population of the excited state.
According to Eq. (\ref{eq3-7}), the population for the excited state $|2\rangle$ is $P_{2}=\sin^{2}{\gamma}$.
The average population of the excited state is thus given as
\begin{align}\label{eq3-13a}
  \overline{P}_{2}=\frac{1}{\gamma_{max}-\gamma_{min}}\int_{\gamma_{min}}^{\gamma_{max}}\sin^{2}{\gamma}d\gamma,
\end{align}
where $\gamma_{max}$ is the maximum value and $\gamma_{min}$
is the minimum value for $\gamma$, respectively.
With $\gamma$ given inform of Eq. (\ref{eqS22}), we have $\gamma_{max}=\gamma_{0}\pi$, $\gamma_{min}=0$, and
\begin{align}\label{eq3-13b}
  \overline{P}_{2}=\frac{1}{2}-\frac{\sin(2\gamma_{0}\pi)}{4\gamma_{0}\pi}.
\end{align}
We use a function $\epsilon=\Gamma_{a}\cdot \overline{P}_{2}\cdot(T\tilde{\Omega}_{0}^{max})$
($\Gamma_{a}$ can be regarded as the average loss rate of the excited state
which relates to the spontaneous emission) to describe the probability
of loss of the excited state. Beware that $\epsilon$ is not limited by 1, so,
in general, we can not directly treat it as the fidelity error caused by spontaneous emission.
When $\tau=0.115T$ and $\tau_{c}=0.3T$, we find
\begin{align}\label{eq3-13c}
  \tilde{\Omega}_{0}^{max}=\tilde{\Omega}_{0}|_{t=0}=\frac{\pi}{\tau}\sqrt{\cot^{2}{(\pi\gamma_{0})}+\frac{4\cot^2{2\varphi}}{\cos^{2}(\pi\gamma_{0})}}.
\end{align}
According to Eqs. (\ref{eq3-13b}) and (\ref{eq3-13c}), we have
$\frac{\partial}{\partial\varphi}\epsilon<0$ ($0<\varphi<\pi/4$)
and $\frac{\partial}{\partial\gamma_{0}}\epsilon<0$ ($0.05<\gamma_{0}<0.3$).
$\epsilon$ is in inverse proportion to both $\gamma_{0}$ and $\varphi$.
Fig. \ref{figS4} which shows $\epsilon$ versus $\gamma_{0}$ and $\varphi$
also verifies this point.
This result means, a large $\gamma_{0}$ which significantly shortens the interaction time
and reduces the total pulse area, however, fails to restrain the influence of spontaneous emission.
This is because when the interaction time is short enough,
further reducing the interaction time can not counteract the negative effect caused
by the excited-state population $P_{2}$. When the interaction time is short enough, reducing the
excited-state population is more important to restrain the influence of spontaneous emission.
One of the effective ways to reduce the
excited-state population, as we know, is increasing the detuning. Nevertheless,
increasing the detuning will inevitably increase the total interaction time.
This is why a large detuning also fails to restrain the influence of spontaneous emission as shown in Fig. \ref{figS4}.
Therefore, taking the requirements of short interaction time, less energy cost,
and robustness against spontaneous emission all into account, an off-resonant system with small detuning
might the best choice to realize the speedup STIRAP in practice.
In the following, we would like to numerically show the influence of spontaneous emission on the fidelity of population transfer
under Markov approximation with Lindblad equation \cite{Cmp48119}
\begin{align}\label{eq3-14}
  \frac{d\rho}{dt}=\frac{1}{i\hbar}[\tilde{H},\rho]+\sum_{n=1,3}{\Gamma_{n}}[S_{n}^{-}\rho S_{n}^{+}-\frac{1}{2}(S_{n}^{+}S_{n}^{-}\rho+\rho S_{n}^{+}S_{n}^{-})],
\end{align}
where $\rho$ is the density matrix, $S_{n}^{-}=|n\rangle\langle 2|$, and $S_{n}^{+}=(S_{n}^{-})^{\dag}$ are
the so-called Lindblad operators, $\Gamma_{n}$ is the spontaneous emission rate from
the excited state $|2\rangle$ to the ground state $|n\rangle$. We plot the fidelity of the population transfer
with considering the small detuning case (choosing $\varphi=\pi/5$) versus $\Gamma_{n}$ in Fig. \ref{fig4}.
The spontaneous emission from the excited state $|2\rangle$ to the ground state $|1\rangle$
affects the fidelity more seriously than that to the ground state $|3\rangle$
as shown in the figure. According to the result, we verify that the population transfer
is robust against spontaneous emission since it is still about $85\%$ even when $\Gamma_{1}=\Gamma_{3}=0.5\tilde{\Omega}_{0}^{max}$.
Here we can find an interesting phenomenon from the figure that
when $\Gamma_{1}$ is relatively large ($\Gamma_{1}>0.2\Omega_{0}^{max}$),
spontaneous emission from the excited state $|2\rangle$ to the ground state $|3\rangle$
becomes a favourable factor in obtaining a high fidelity of the scheme.
The reason behind this phenomenon is because the state $|3\rangle$ is just the target state,
so spontaneous emission from the excited state $|2\rangle$ to the ground state $|3\rangle$ indeed
increases the fidelity.

Our scheme can be realized with the setup of Ref. \cite{arxiv1607}, where a solid-state $\Lambda$
system is hosted by a single nitrogen-vacancy centre in diamond at low temperature,
and the shapes and phases of the driving pulses can be modulated by electrooptic modulators. With a single
tunable laser (637.2nm), the maximal amplitude of the modified
pulses can be chosen as $\tilde{\Omega}_{0}=2\pi\times171$MHz (under the RWA),
the corresponding operation time is $T\approx14$ns which is smaller than
the life time for the $|A_{2}\rangle=|2\rangle$ spin-orbit excited state at temperature $5.5$K.
This result is very close to that in Ref. \cite{arxiv1607} with resonant driving system, and it is much better than that (maximal amplitude of the modified
pulses is $\tilde{\Omega}_{0}\approx6\times 2\pi$MHz and operation time $T=0.4$ms)
in Ref. \cite{Natc712479} with highly detuned pulses. Ref. \cite{arxiv1607} also reports the relaxation rates
$\Gamma_{1}/2\pi=4.3$MHz and $\Gamma_{3}/2\pi=8.5$MHz for decay of the
excited $|A_{2}\rangle$ level into the $|1\rangle$ and $|3\rangle$ ground states, respectively.
With the experimental data, we can find the fidelity of the population transfer is $97.48\%$.

\section{Conclusion}
We have shown that
the intermediate Hamiltonian could be directly designed by using the original Hamiltonian
and the CD Hamiltonian.
With the help of the easy-to-get intermediate Hamiltonian,
we have presented a general approach to
remove the adiabatic condition to speed up
the corresponding population transfer process
without adding the traditional CD Hamiltonian.
Our method is significantly different from that of Ref. \cite{Prl116230503}, where the important intermediate Hamiltonian is constructed
with some specified dressed states.
The present approach is physically transparent
and extremely flexible.
We further show
with the help of the intermediate Hamiltonian, speedup scheme
can be realized even without breaking down the structure
of the original Hamiltonian, which is important in view of experiment.

Further applications or extensions of this approach could be in
fields such as shortcuts to non-Hermitian quantum adiabatic computation \cite{Pra84023415,Pra93052109,Pra8705250289063412},
dissipative master equations \cite{Pra78033417,Njp16053017}, or more complicated systems \cite{Pra94043623,Njp18012001,Pra95022332}.

This work was supported by the National Natural Science Foundation
of China under Grants No. 11575045, No. 11374054 and No. 11675046.


\end{document}